\documentclass[preprint,12pt]{elsarticle}

\usepackage{natbib}
\biboptions{numbers,super,sort&compress}

\usepackage{amssymb}
\usepackage{amsmath}
\usepackage[hidelinks]{hyperref}

\journal{IJRMMS}
\myfooter[C]{\parbox{0.95\textwidth}{\centering
Accepted manuscript\\
Published in International Journal of Rock Mechanics and Mining Sciences (2026). \\ DOI: 10.1016/j.ijrmms.2026.106431\\
Licensed under CC BY-NC-ND 4.0 (https://creativecommons.org/licenses/by-nc-nd/4.0/).}}

\begin{document}

\begin{frontmatter}



% \title{Identifying dissolution regimes and their impact on heterogeneity in porous and fractured rocks}
\title{Structural barriers to complete homogenization and wormholing in dissolving porous and fractured rocks}

\author[label1]{Tomasz Szawełło\corref{cor1}}
\ead{t.szawello@uw.edu.pl}
\author[label2,label3]{Jeffrey D. Hyman}
\author[label4,label5]{Peter K. Kang}
\author[label1]{Piotr Szymczak\corref{cor1}}
\ead{piotrek@fuw.edu.pl}
\affiliation[label1]{organization={Institute of Theoretical Physics, Faculty of Physics, University of Warsaw},
            city={Warsaw},
            postcode={02-093},
            country={Poland}}

\affiliation[label2]{organization={Los Alamos National Laboratory, Earth and Environmental Sciences Division},
            city={Los Alamos},
            postcode={87545},
            state={NM},
            country={USA}}

\affiliation[label3]{organization={Department of Geology and Geological Engineering, Colorado School of Mines},
            city={Golden},
            postcode={80401},
            state={CO},
            country={USA}}

\affiliation[label4]{organization={Department of Earth \& Environmental Sciences, University of Minnesota},
            city={Minneapolis},
            postcode={55455},
            state={MN},
            country={USA}}

\affiliation[label5]{organization={Saint Anthony Falls Laboratory, University of Minnesota},
            city={Minneapolis},
            postcode={55455},
            state={MN},
            country={USA}}

\cortext[cor1]{Corresponding author.}
%% Abstract
\begin{abstract}

Dissolution in porous media and fractured rocks alters both the chemical composition of the fluid and the physical properties of the solid. Depending on system conditions, reactive flow may enlarge pores uniformly, widen pre-existing channels, or trigger instabilities that form wormholes. The resulting pattern reflects feedbacks among advection, diffusion, surface reaction, and the initial heterogeneity of the medium. Porous and fractured media can exhibit distinct characteristics—for example, the presence of large fractures can significantly alter the network topology and overall connectivity of the system.
% We quantify these differences using network models representing a regular porous medium, a disordered porous medium, and a discrete fracture network, all analyzed with a unified metric—the flow focusing profile.
We quantify these differences with three network models---a regular pore network, a disordered pore network, and a discrete fracture network---evaluated with a unified metric: the flow focusing profile.
This metric effectively captures evolution of flow paths across all systems: it reveals a focusing front that propagates from the inlet in the wormholing regime, a system-wide decrease in focusing during uniform dissolution, and the progressive enlargement of pre-existing flow paths in the channeling regime. The metric shows that uniform dissolution cannot eliminate heterogeneity resulting from the network topology. This structural heterogeneity---rather than just pore-diameter or fracture-aperture variance---sets a fundamental limit on flow homogenization and must be accounted for when upscaling dissolution kinetics from pore or fracture scale to the reservoir level. 

\end{abstract}



% \item Wormholing, channeling, and uniform dissolution arise in regular porous media, disordered porous media, and discrete fracture networks alike; however, the characteristics of the regimes are different, shaped by the inherent heterogeneity of each system.

% \item Uniform dissolution can eliminate heterogeneity associated with pore size or aperture variance, but it cannot erase heterogeneity that stems from the network topology---specifically, differences in connectivity or in the lengths of individual flow pathways.
% \begin{highlights}

% \item Dissolution regimes in porous and fractured rocks are shaped by system heterogeneity 

% \item Dissolution is able to remove pore size or aperture-related heterogeneity

% \item Heterogeneity from connectivity or length variability persists after dissolution
% \end{highlights}

%% Keywords
\begin{keyword}
Pore Network Model \sep Discrete Fracture Network \sep Flow Focusing Profile \sep Dissolution Regime \sep Heterogeneity


\end{keyword}

\end{frontmatter}



\section{Introduction}
\label{sec1}
%% Labels are used to cross-reference an item using \ref command.
The flow of reactive fluids through rock induces significant changes in both the fluid composition and the properties of the solid matrix, thereby affecting the transport dynamics within the system. Understanding these processes is essential for a wide range of geological and environmental applications. For instance, the safe disposal of high-level radioactive waste depends on the ability to predict the dominant modes of transport in the reservoir.~\cite{tsang2015, churakov2020} Likewise, assessing contaminant migration requires understanding how flow is focused, because residence time determines whether a pollutant can be neutralized.~\cite{fryar1998, kung2000, mayer2001} In both contexts, transport is tightly coupled to the evolving pore structure: reactive fluid remodels the medium and creates preferential paths that ultimately control system behavior.~\cite{berkowitz1998, kang2015, hyman2016} In the context of $\textrm{CO}_2$ sequestration, understanding both micro- and macro-scale behavior is critical for process optimization.~\cite{andreani2009, matter2009, steefel2013, hyman2020, nisbet2024} Predicting the evolution of flow paths enables the avoidance of clogging and maximization of mineral replacement.~\cite{budek2025} It is therefore evident that elucidating the coupling between flow, transport, and the evolving properties of the reservoir is essential to explain the diversity of structures that emerge from dissolution.~\cite{chadam1986, daccord1987a, hoefner1988, golfier2002}

A wide body of experimental and numerical work has quantified how dissolution patterns depend on the relative time scales of advection, diffusion, and reaction. In porous media, dissolution experiments have investigated the pore scale behavior~\cite{molins2014, menke2015, dutka2020} and patterns emerging at the core scale.~\cite{hoefner1988, golfier2002, fredd1998, garcia-rios2017, ott2015, snippe2020, cooper2023} Their findings have been reproduced and extended by multi-scale simulations, from direct numerical models to continuum codes.~\cite{hoefner1988, molins2014, panga2005, szymczak2009, hao2013, steefel2015, budek2012, menke2023} Four principal dissolution regimes are now recognized: compact, wormholing, channeling, and uniform. In compact (face) dissolution, the reactant is exhausted almost immediately after entering the sample, and a reaction front advances uniformly across its width.~\cite{hoefner1988, cohen2008} When wormholing occurs, advective instabilities localize flow into a few highly conductive channels that propagate from inlet to outlet.~\cite{szymczak2009, ortoleva1987, mcduff2010} In uniform dissolution, the reactant penetrates the full length of the system, enlarging pores homogeneously.~\cite{golfier2002, roded2020} Finally, in the channeling regime, initially existing flow paths are uniformly enlarged along their entire length.~\cite{menke2023, menke2016} Because the transitions between these regimes are subtle, distinguishing between them requires quantitative analysis of the system evolution.~\cite{jang2011, szawello2024}

Analogous dissolution phenomena also occur in fractured media, where high-permeability discontinuities channel most of the flow, distinguishing them from conventional porous media. Studies of dissolution in single-fracture systems—using both numerical models and flow-through experiments—have extensively documented aperture growth and channel development.~\cite{garcia-rios2017, dreybrodt1990a, dreybrodt1996,  durham2001, detwiler2003, li2008, szymczak2004} Similarly to porous media, studies of two-dimensional fracture planes have revealed compact, wormhole, and uniform dissolution regimes, each governed by the interplay of flow, transport, and chemical reactions.~\cite{szymczak2009,hanna1998,detwiler2007} Although the coupled flow–transport–reaction equations can be solved directly for idealized one- or two-dimensional fractures, doing so for natural systems with thousands of intersecting fractures—such as karst conduit networks—remains computationally formidable. To address this challenge, researchers employ a hierarchy of different representations: two-dimensional grids of intersecting fractures,~\cite{dreybrodt1996,gabrovsek2001,dreybrodt2005} fully three-dimensional networks with spatially variable apertures,~\cite{li2020,jiang2022,jiang2023} and stochastic discrete fracture network (DFN) models.~\cite{cacas1990,hyman2015} DFNs are attractive because they incorporate field-derived statistics of fracture aperture, length, and orientation, bridging single-fracture physics and network-scale heterogeneity while remaining computationally tractable for reactive transport simulations.~\cite{pachalieva2025}

Porous media and fractured media exhibit fundamentally different physical structures and flow behaviors---fractures, for instance, can generate long-range, highly connected flow paths---yet the impact of these distinct forms of heterogeneity on dissolution dynamics remains poorly understood and has not been systematically compared.
Early work on variable-aperture fractures showed that the volume of reactant required for breakthrough varies non-monotonically with the magnitude of the initial aperture variation; a critical level of heterogeneity minimizes the breakthrough volume.~\cite{hanna1998, cheung2002} The same trend was reproduced in numerical analysis of porous media that imposed spatially correlated porosity fields.~\cite{kalia2009, maheshwari2013a, maheshwari2013} Follow-up studies examined the role of heterogeneity length scale~\cite{maheshwari2013} and suggested that the strong heterogeneity–wormholing link observed in small domains diminishes in larger systems.~\cite{upadhyay2015} There is, however, agreement that heterogeneity controls competition among channels and promotes tip branching during wormhole growth.~\cite{cheung2002, upadhyay2015}

The interplay between heterogeneity and evolving dissolution patterns strongly affects the hydraulic properties of a medium. Pore network simulations show that uniform dissolution can eliminate all variability in pore diameters,~\cite{roded2020, deng2025} whereas dissolving an initially homogeneous medium can instead generate pronounced heterogeneity; these opposing outcomes, in turn, remove or create anomalous transport behavior.~\cite{berkowitz1998, deng2025, saeibehrouzi2025} Heterogeneity in porous media has also been shown to affect the onset of preferential flow,~\cite{cooper2023, menke2016, al-khulaifi2019} permeability evolution,~\cite{li2019} mixing,~\cite{sharma2023} and even the overall dissolution regime.~\cite{menke2023, szawello2024, kanavas2025} These results show that significant changes in system dynamics can arise solely from adjusting pore-diameter variability. Beyond pore-size variability, heterogeneity also stems from the topology and geometry of pore or fracture networks—specifically, differences in connectivity and in segment-length distributions. This structural heterogeneity also shapes system evolution, influencing processes ranging from karst genesis~\cite{aliouache2024} to mineral carbonation.~\cite{hyman2024}

In this paper, we utilize the flow focusing metric introduced by Szawełło~et~al.~\cite{szawello2024} to quantify how different forms of heterogeneity couple with dissolution across a range of flow and reaction conditions. We apply the metric to three network models: a regular pore network with variance in pore diameters, a disordered pore network with variance in pore diameters and lengths, and a discrete fracture network with variance in fracture apertures, lengths, and connectivity. We distinguish three heterogeneity types: conduit-scale (pore diameter or fracture aperture), segment-scale (length), and network-scale (connectivity). By tracking each scale through time, we show that initial heterogeneity shapes the evolution of flow paths and ultimately limits the degree to which dissolution can homogenize the system. Our focus is on the impact of length and connectivity, which together form a quenched disorder in the network—that is, a structural heterogeneity that dissolution can modify only marginally. Initially, this disorder plays a smaller role than the widely studied heterogeneity in conduit width: pore diameters enter the conductance in the fourth power and fracture apertures in the third, whereas lengths enter only linearly. As homogenization progresses and variation in conduit widths diminishes, however, segment- and network-scale heterogeneity become the decisive factors in flow-path selection. This behavior is key to understanding the differing responses of porous media and fracture networks to dissolution.

%Tracking the evolution---or persistence---of each type reveals that all three model systems can enter the same dissolution regimes---uniform, channeling, and wormholing---yet the original heterogeneity shapes the characteristics of each regime. Most notably, it governs the extent to which dissolution can homogenize the flow pathways.

\section{Reactive Transport in Network Models}
We quantitatively describe the dissolution regimes and analyze their transport properties by utilizing network models. We use a capillary pore network model for the dissolving porous medium~\cite{hoefner1988, budek2012} and a graph representation of a discrete fracture network for dissolving fractures.~\cite{cacas1990, hyman2015} We adjust the initial properties, that is, connectivity, conduit lengths, and distributions of pore diameters and fracture apertures, to exhibit the differences in dissolution arising from the structure of the medium.

\subsection{Network Generation}
\begin{figure}
\centering
\noindent\includegraphics[width=1.0\textwidth]{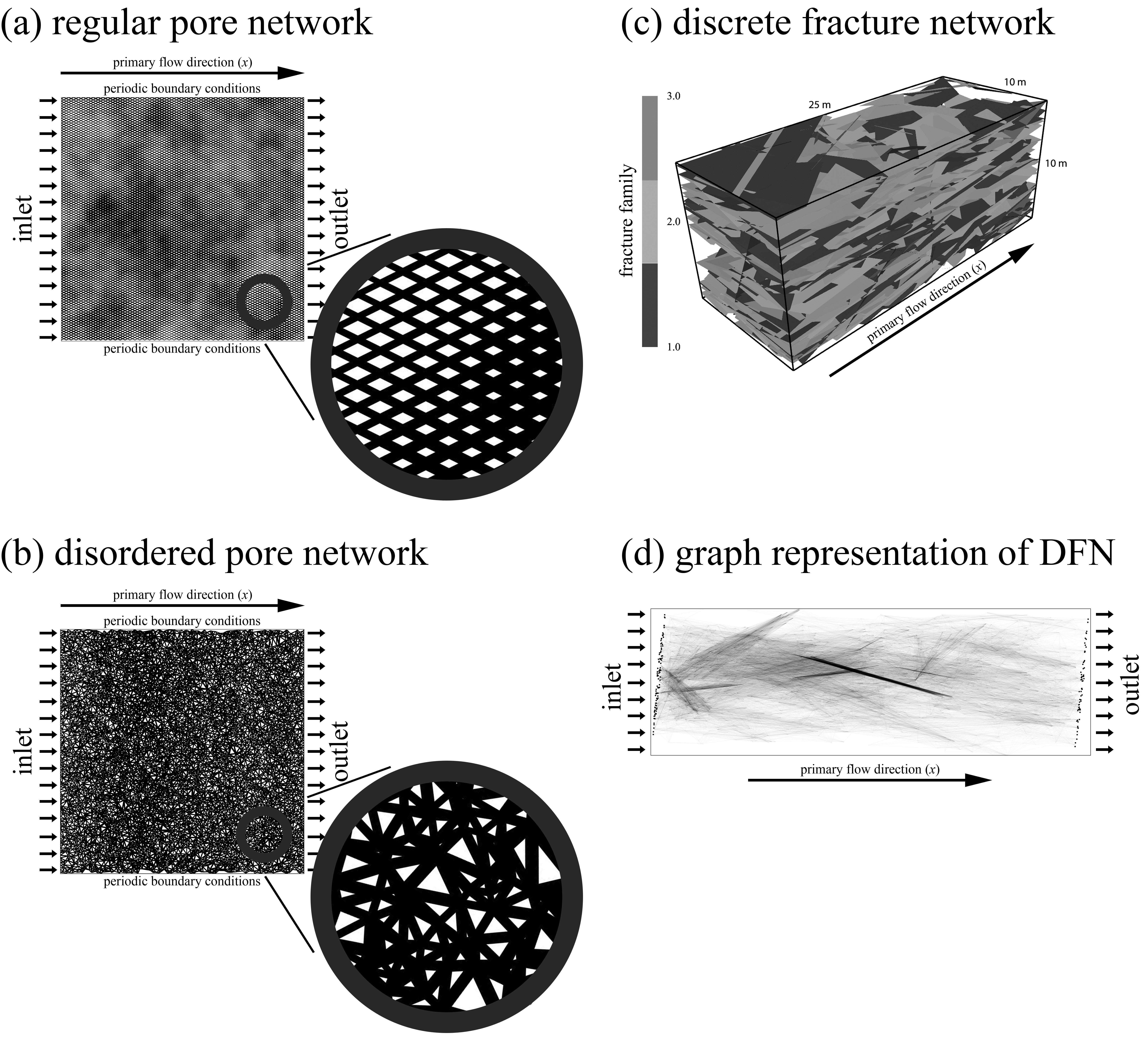}
    \caption{(a) Example regular pore network realization (diamond lattice); edge width is proportional to the initial pore diameter. (b) Example disordered pore network realization (Delaunay lattice); edge width again reflects the initial pore diameter. (c) Example discrete fracture network; colors denote fracture families. (d) Graph representation of the same DFN, projected into two dimensions via principal component analysis. Edge width is kept constant, so color intensity corresponds to the local density of fracture segments.}
    \label{fig:network}
\end{figure}

We investigate dissolution dynamics in three contrasting network geometries.
\begin{enumerate}
    \item Regular pore network: a diamond lattice in which every edge has identical length; variability arises solely from the log-normal distribution of pore diameters (conduit-scale heterogeneity).
    \item Disordered pore network: a Delaunay network whose nodes are randomly positioned, introducing heterogeneity in pore lengths and intersection angles while retaining the same diameter distribution as the regular lattice (conduit- and segment-scale heterogeneity).
    \item Discrete fracture network: a semi-generic model loosely based on the fractured carbonate-hosted Pietrasecca Fault in the central Apennines, Italy~\cite{smeraglia2021}; node locations, fracture intensities, and aperture statistics reproduce the multiscale heterogeneity of natural carbonate fractures (conduit-, segment-, and network-scale heterogeneity).
\end{enumerate}
Fig.~\ref{fig:network} shows a representative example of each network.

For every geometry, we generate 30 statistically equivalent realizations. In the two pore network cases, we use a domain of side length $L$ discretized into $100\times100$ nodes. Periodic boundary conditions are applied in the direction perpendicular to the imposed pressure gradient: pores on one side of the lattice are connected to their counterparts on the opposite side, which removes artificial side-wall effects.

For the regular pore network, each realization shares the same connectivity, but draws pore diameters from an independent log-normal distribution with mean $d_{0}$, variance $d_0^2$ (that is, the coefficient of variation is $\sim 1$), and correlation length of $0.1L$; the sampling procedure follows Szawełło~et~al.~\cite{szawello2024} and Upadhyay~et~al.~\cite{upadhyay2015}
The value of $d_0$ can range from micrometers to millimeters, depending on the mineral and rock type; because the model is formulated in dimensionless variables, its exact value is not prescribed. The formulation is, however, most appropriate for long, slender pores, so that dissolution does not drastically alter the network topology. In the regular network, all pore lengths are set equal to a reference length $l_0$.

For the disordered pore network, nodes are randomly distributed in the domain and connected using a Delaunay triangulation; node positions differ between realizations.
%To impose periodic boundary conditions, we tile the domain with its translated copies, perform the Delaunay triangulation on this extended point set, and then retain only those edges whose midpoints lie inside the original domain, identifying edges that cross a lateral boundary with periodic connections on the opposite side.
To impose periodic boundary conditions, we tile the original domain with its translated copies and construct a Delaunay triangulation on this enlarged set of points. From the resulting edges, we keep only those whose midpoints lie inside the original domain. For edges that cross a side boundary, we connect them to the matching points on the opposite side, creating periodic links.
Pore diameters are assigned from the same distributions as for the regular network, while pore lengths are given by the Euclidean distance between connected nodes; by construction, the average pore length is equal to $l_0$.

The DFN domain (25 m × 10 m × 10 m) is populated with three fracture families that differ in orientation, length distribution, aperture statistics, and intensity; full generation parameters are listed in~\ref{appendix:dfngeneration}. Importantly, each fracture family has a set aperture, constant throughout the realizations, but the proportions between families vary. We convert each DFN to a pipe-network graph with the algorithm of Hyman~et~al.,~\cite{hyman2018} preserving fracture intersections as graph nodes.

In every model, the nodes on one boundary serve as the inlet, while those on the opposite boundary form the outlet.

\subsection{Reactive Transport}
Our reactive transport model follows the derivations of Budek~and~Szymczak~\cite{budek2012} for pore networks and Szymczak~and~Ladd~\cite{szymczak2012} for single fractures. In the pore network representations (Fig.~\ref{fig:network}a,b), each edge corresponds to a cylindrical pore of diameter $d$ and length $l$; in the discrete fracture network (Fig.~\ref{fig:network}c,d), each edge represents a fracture segment of aperture $b$ and length $l$ spanning a fracture of width $w$. Volumetric flow $q$ in each channel obeys
\begin{equation}
    q = -\frac{C}{\mu} \nabla P,
\end{equation}
where $C$ is the hydraulic conductance, $P$ the pressure, and $\mu$ the fluid viscosity. For a cylindrical pore $C = \pi d^4 / 128$ (Hagen–Poiseuille equation); for a fracture slot $C = w b^3/12$ (Reynolds equation). At each intersection $i$, we impose mass conservation in the form of the nodal continuity condition
\begin{equation}
    \sum_j q_{ij} = 0,
\end{equation}
where the sum is over all edges $ij$ incident on node $i$. Across the network, a constant total volumetric flow rate $Q$ is enforced by setting $P=P_0$ at the inlet nodes and $P=0$ at the outlet, then rescaling $P_0$ at every time step to maintain $Q$.

We consider dissolution of a porous medium or a fracture network by an aqueous reactant of inlet concentration $c_\textrm{in}$, injected at the inlet nodes. We assume a single-component dissolution reaction with a linear rate law
\begin{equation}
    R(c_\textrm{w}) = k c_\textrm{w},
    \label{eq:rate}
\end{equation}
where $k$ is the surface reaction rate and $c_\textrm{w}$ is the reactant concentration at the pore/fracture wall. This rate law is a standard approximation for far-from-equilibrium reactions, such as dissolution by concentrated acid; in that case $c_\textrm{w}$ corresponds to the concentration of $\textrm{H}^+$ ions.~\cite{hoefner1988} An equivalent formulation appears in the karst literature, where limestone dissolution is often treated as first order in undersaturation, which maps directly onto our driving concentration variable.~\cite{dreybrodt1990a}

It is convenient to express the wall concentration in terms of the bulk concentration $c$. The rate law in Eq.~\eqref{eq:rate} can then be written as $R(c) = k_\textrm{eff}(d^\textrm{h}) c$, where $k_\textrm{eff}(d^\textrm{h})$ is an effective reaction rate that incorporates the hindering effects of transverse diffusion:
\begin{equation}
    k_\textrm{eff}(d^\textrm{h})=\frac{k}{1+\frac{kd^\textrm{h}}{D \textrm{Sh}}},
    \label{eq:keff}
\end{equation}
 where $D$ is the diffusion coefficient, $\textrm{Sh}$ is the Sherwood number,~\cite{bejan1984, hayes1994, gupta2001} and $d^\textrm{h}$ is the hydraulic diameter, equal to $d$ for pores and $2b$ for fractures. The Sherwood number, Sh, depends on reaction rate at mineral surfaces, but the variation is relatively small, for porous media limited by the values $\textrm{Sh} = 3.656$ and $\textrm{Sh} = 4.364$ (we approximate it by a constant value, $\textrm{Sh} = 4$), while in fractured media, from $\textrm{Sh} = 7.54$ to $\textrm{Sh} = 8.24$ (we approximate it by $\textrm{Sh} = 8$).

Along each conduit we solve the one-dimensional advection–reaction equation for the bulk concentration
\begin{equation}
	q \frac{dc}{dx} = - s k_\textrm{eff}(d^\textrm{h}) c,
    \label{eq:transport}
\end{equation}
where $x$ is the coordinate along the channel length, the reactive surface is $s = \pi d$ for pores and $s = 2 w$ for fractures.

While in Eq.~\eqref{eq:transport} the effective reaction rate $k_\textrm{eff}(d^\textrm{h})$ incorporates the effects of transverse diffusion, axial diffusion (along the flow direction) is neglected. In general, this approximation is justified when the P\'eclet number, which measures the relative importance of advection and diffusion, is larger than one. The definition of the P\'eclet number requires a choice of length scale; for reactive infiltration this scale is set by the penetration length of the reactant, that is, the characteristic distance it can travel into the medium before being consumed. If, on this length scale, advection dominates over axial diffusion, then axial diffusion can be safely neglected.

At each intersection $i$, we assume perfect mixing of the concentration,
\begin{equation}
    c_i = \frac{\sum_{j^\prime} q_{ij} c_{ij}^\textrm{out}}{\sum_{j^\prime} q_{ij}},
\end{equation}
where the sum runs over all edges with flow directed into node $i$ and $c_{ij}^\textrm{out}$ is the outlet concentration at the end of those edges. The node concentration $c_i$ is then used as the inlet concentration for all edges with flow directed out of node $i$.

Mineral dissolution enlarges each conduit according to
\begin{equation}
    \partial_t d = \frac{2 k_\textrm{eff}(d)}{\nu c_\textrm{sol}}\, c, \qquad
    \partial_t b = \frac{2 k_\textrm{eff}(2b)}{\nu c_\textrm{sol}}\, c,
\end{equation}
where $c_\textrm{sol}$ is the molar concentration of soluble mineral in the solid and $\nu$ is the stoichiometric coefficient of the reaction. To keep the model tractable, in each time step we compute the total volume of mineral dissolved along the conduit and increase its diameter (or aperture) uniformly so that the conduit volume increases by exactly that amount, thereby preserving mass conservation.

The model assumes a separation of time scales between transport relaxation and mineral dissolution: for each geometry we solve the steady-state flow and transport problem (Eq.~\eqref{eq:transport}) and then update $d$ or $b$ using the above law. This quasi-steady approximation is appropriate when the acid capacity number $\gamma = c_\textrm{in}/(\nu c_\textrm{sol})$, which measures the volume of solid dissolved by a unit volume of reactant, is much smaller than one. This condition is satisfied in the great majority of geological and industrial reactive transport processes. There are, however, systems where this approximation can break down, for example in the dissolution of halite or caramel.~\cite{cohen2020}

System evolution is governed by two dimensionless numbers: the effective Damk\"ohler number, $\textrm{Da}_\textrm{eff}$, measuring the ratio of advective to reactive time scales, and the reaction-diffusion parameter $\textrm{G}$, characterizing the hindering effects related to diffusion across the channel.
\begin{equation}
    \textrm{Da}_\textrm{eff} = \frac{s_0 k_\textrm{eff}(d^{\textrm{h}}_0) L}{Q}, \ \ \textrm{G} = \frac{k d^{\textrm{h}}_0}{D \textrm{Sh}},
\end{equation}
where $s_0$ and $d^{\textrm{h}}_0$ are the initial average reactive surface and hydraulic diameter, respectively.

The two dimensionless groups we employ—the effective Damköhler number $\mathrm{Da}_{\mathrm{eff}}$ and the transport ratio $\textrm{G}$—quantify the competition of advective, diffusive, and reactive time scales in both porous and fractured media.  We define them with the total volumetric flow rate~$Q$
rather than a mean velocity, because strong initial flow focusing in discrete fracture networks (DFNs) makes any ``average'' velocity
unrepresentative.  This volumetric definition also remains consistent for our pore network simulations, which share identical width and an equal number of inlet pores, so $Q$ scales directly with the mean pore velocity.

A direct, one–to–one comparison of $\mathrm{Da}_{\mathrm{eff}}$ between the two systems is nevertheless impossible.  In our models, a cross-section perpendicular to the main flow contains only 200 conduits in the pore networks but roughly 5000 fracture elements in the DFNs; the vastly different hydraulic cross-sections imply different residence-time distributions even at the same global flow rate~$Q$.  Consequently, we treat $\mathrm{Da}_{\mathrm{eff}}$ as a system-specific control parameter, scanning it separately for pore and fracture networks and then comparing the resulting dissolution regimes rather than matching absolute values.

Simulation time is expressed as a dimensionless dissolved volume,
\begin{equation}
    T = \frac{V_\textrm{diss}}{V^\textrm{p}_0},
\end{equation}
where $V^\textrm{p}_0$ is the initial total pore (or fracture) volume and $V_\textrm{diss}$ is the cumulative volume of solid that has dissolved. Thus, $T$ measures the total dissolved solid volume, normalized by the initial pore volume.

The above model is implemented in our in-house network simulation code, written in Python, as described in Szawełło,~\cite{szawello2025_soft} where the source code and input files are openly available.

\subsection{Model Assumptions and Limitations}
The model relies on several simplifying assumptions and approximations, which we summarize here for clarity. First, we represent the porous medium as a network of cylindrical pipes and the fractured medium as a network of rectangular fracture segments. Both the geometry of each conduit and the intersections (which we treat as volumeless nodes) are simplified in this way. This idealization is standard in pore network and DFN models and is what allows us to extend modeling beyond the scale of individual pores while retaining analytical expressions for flow and transport within each conduit. 
% In this study we make one additional assumption about the conduit geometry: conduits do not merge. Since our main interest is in homogenization at the network scale, we avoid topological changes that would create highly localized flow focusing in newly merged pores compared with their still-separate neighbors. Physically, this is a reasonable approximation for fractured media, in which fracture apertures and the distances between fractures can vary by orders of magnitude. In porous media, it corresponds to long, slender pores and, more generally, low-porosity rocks, where neighboring pores remain distinct even under substantial dissolution.

A second approximation concerns how dissolution modifies the conduit geometry. Because the reactant concentration decays along each conduit, the local dissolution rate is largest near the upstream end and smallest near the downstream end. In our model, we compute the total volume of mineral dissolved along the conduit and increase its diameter (or aperture) uniformly. This update is most accurate when the reactant penetration length,
\begin{equation}
    l_\textrm{p} = \frac{L}{\textrm{Da}_\textrm{eff}},
\end{equation}
is large compared with the average conduit length, $l_\textrm{p} \gg l_0$. The same scale enters the treatment of axial diffusion: neglecting diffusion along the flow direction is justified when the Péclet number based on $l_\textrm{p}$ is greater than one, so that advection dominates on the scale over which the reactant is consumed. This restriction means that the model is not intended to describe strongly diffusion-dominated regimes such as compact dissolution.

Additional assumptions are made regarding transport and chemical reactions. We assume a separation of time scales between flow and transport relaxation and mineral dissolution (acid capacity number $\gamma \ll 1$). We also assume perfect mixing of reactant at each intersection, which is a reasonable approximation for disordered networks but may overestimate mixing in simple regular lattices at high P\'eclet numbers, where streamline-based mixing models can be more appropriate.~\cite{sharma2023} Finally, the chemical kinetics are represented by a single-component linear rate law. This is a common approximation across a range of systems, but care is needed when applying it to systems with multiple reactive species and strong buffering, where equilibrium reactions can substantially modify the fluid chemistry (for example by buffering the $\textrm{H}^+$ concentration).

\subsection{Flow Focusing Profile}
The main tool that we use to determine the evolution of the system is the flow focusing profile.~\cite{szawello2024} We segment the medium into cross sections along the main flow direction, $x$, and in each of them we calculate the flow focusing index, $f_{50\%}$,~\cite{jang2011} according to
\begin{equation}
    f_{50\%} = \frac{n_{\textrm{X}} / 2 - n_{50\%}}{n_{\textrm{X}} / 2},
    \label{eq:flow_focusing_index}
\end{equation}
where $n_{50\%}$ is the smallest number of conduits carrying 50\% of the total flow through a given cross section of the medium, and $n_{\textrm{X}}$ is the total number of conduits in that cross section. Calculating the index, $f_{50\%}$, for cross sections along the entire medium, we obtain a profile measuring flow focusing as a function of distance from the inlet at a given time.

\section{Results}
\begin{figure}
\centering
\noindent\includegraphics[width=1.0\textwidth]{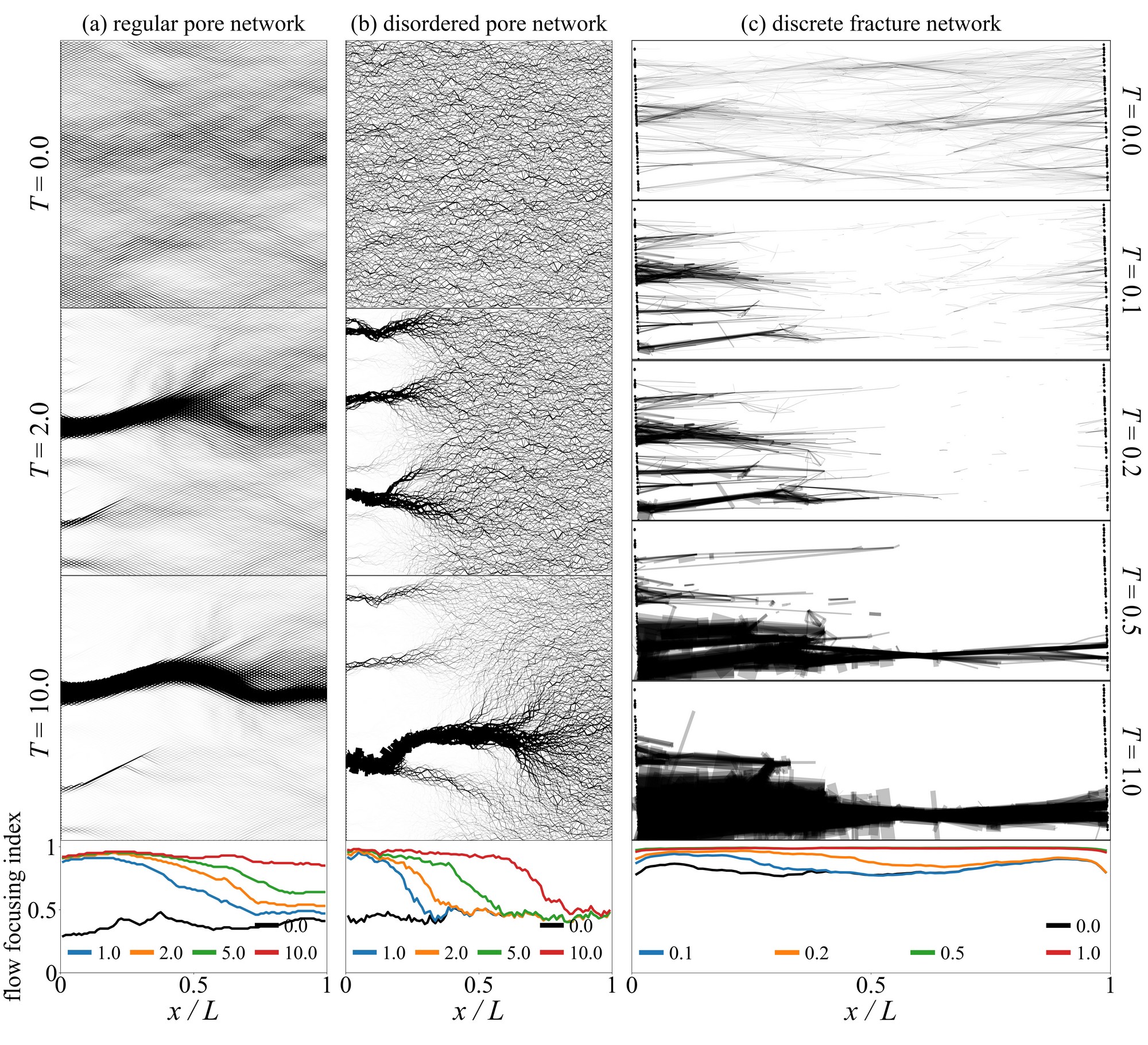}
    \caption{Evolution of the three network types and the flow focusing profile in the wormholing regime. (a) Regular pore network ($\textrm{Da}_\textrm{eff} = 0.2$, $\textrm{G} = 5$). (b) Disordered pore network ($\textrm{Da}_\textrm{eff} = 0.2$, $\textrm{G} = 5$). (c) Discrete fracture network ($\textrm{Da}_\textrm{eff} = 0.02$, $\textrm{G} = 5$) shown as a graph after a principal-component-analysis projection. In each case, edge width is proportional to the volumetric flow rate, with the same proportionality constant used for a given network type. For the DFN, only edges carrying more than 1\% of the maximum flow rate in the system are plotted. In the flow focusing plots, the initial profile is shown by the black line, and profiles at later times ($T = 1.0, 2.0, 5.0,$ and $10.0$ for both pore networks; $T = 0.1, 0.2, 0.5,$ and $1.0$ for the fracture network) are shown by colored lines.}
    \label{fig:wormhole}
\end{figure}
Each network type exhibits distinct dissolution behavior due to differences in structural organization and type of heterogeneity. For the regular pore network, all conduits are of the same length, and only conduit-scale heterogeneity is present, in the form of the distribution of diameters. In the disordered pore network, apart from the conduit-scale heterogeneity, based on the same noise as in the regular network, there is also the segment-scale heterogeneity in the form of distribution of conduit lengths and intersection positions. While in the case of porous media, the conduit- and segment-scale heterogeneity can be of the same order of magnitude, their couplings with dissolution are different, with the latter much more persistent. Finally, in the discrete fracture network, the nodes, edges, and apertures are distributed to approximate a real fracture network, introducing heterogeneity across all scales. Fracture families with various apertures introduce the conduit-scale heterogeneity, variance in fracture lengths generates the segment-scale heterogeneity (which in fractures can be much larger than that in apertures), and finally connectivity within the structure creates heterogeneity on the scale of the entire network, as sometimes only a few fractures can span the entire domain.

We simulate the three systems, choosing the dimensionless parameters $\textrm{Da}_\textrm{eff}$ and $\textrm{G}$ to explore the standard dissolution regimes: uniform dissolution, channeling, and wormholing. Fig.~\ref{fig:wormhole} presents the networks evolving in the wormholing regime, alongside the changes of the flow focusing profile. In pore networks, the dissolution regime becomes apparent almost immediately through the emergence of one or a few dominant channels. In the DFN the pattern is subtler---flow is further concentrated along pre-existing high-conductivity paths---but a newly developed high-permeability flow path can still be discerned. In each case the dissolution dynamics are captured by the flow focusing profile, which appears as a front of increased focusing that advances from the inlet toward the outlet; this increase is less pronounced in the DFN because strong focusing already exists at $T=0.0$.

Because we define the profile in the same way for every network, it acts as a single quantitative metric for both porous and fractured media. Tracking its evolution in ($\textrm{Da}_\textrm{eff}$, $\textrm{G}$) space lets us compare dissolution in DFNs with that in regular and disordered pore networks, exposing both shared regimes and geometry-specific differences.

\subsection{Regular Pore Network}
\begin{figure}
\centering
\noindent\includegraphics[width=1.0\textwidth]{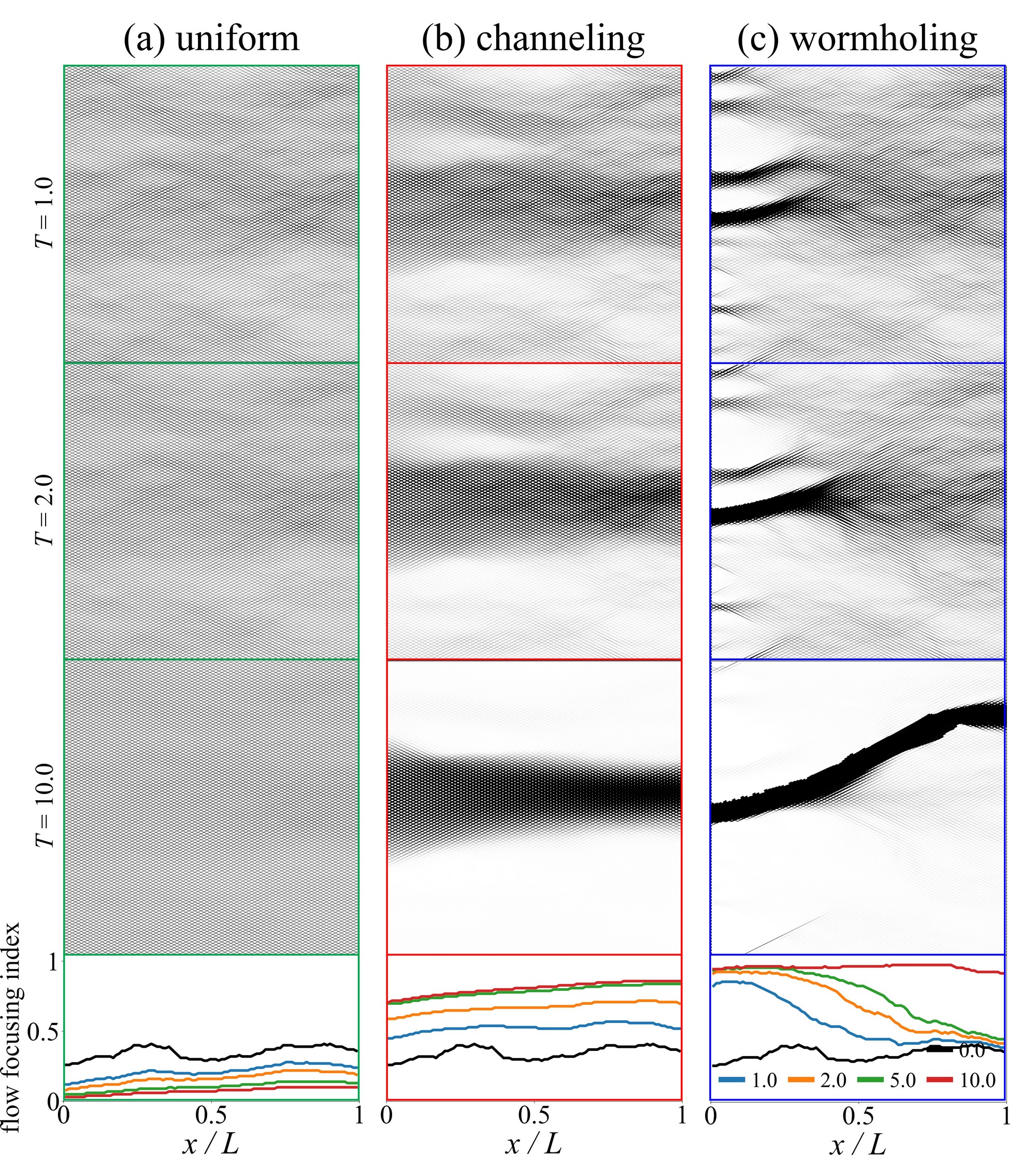}
    \caption{Evolution of the regular pore network and the flow focusing profile in three dissolution regimes: (a) uniform ($\textrm{Da}_\textrm{eff} = 0.002$, $\textrm{G} = 5$), (b) channeling ($\textrm{Da}_\textrm{eff} = 0.02$, $\textrm{G} = 5$), and (c) wormholing ($\textrm{Da}_\textrm{eff} = 0.2$, $\textrm{G} = 5$). Edge width is proportional to the volumetric flow rate, with the same proportionality constant used in all panels. The plots show the initial flow focusing profile (black line) and profiles at times $T = 1.0, 2.0, 5.0,$ and $10.0$ (colored lines). The frame color denotes the dissolution regime: green for uniform, red for channeling, and blue for wormholing.}
    \label{fig:regular_r}
\end{figure}

\begin{figure}
\centering
\noindent\includegraphics[width=0.80\textwidth]{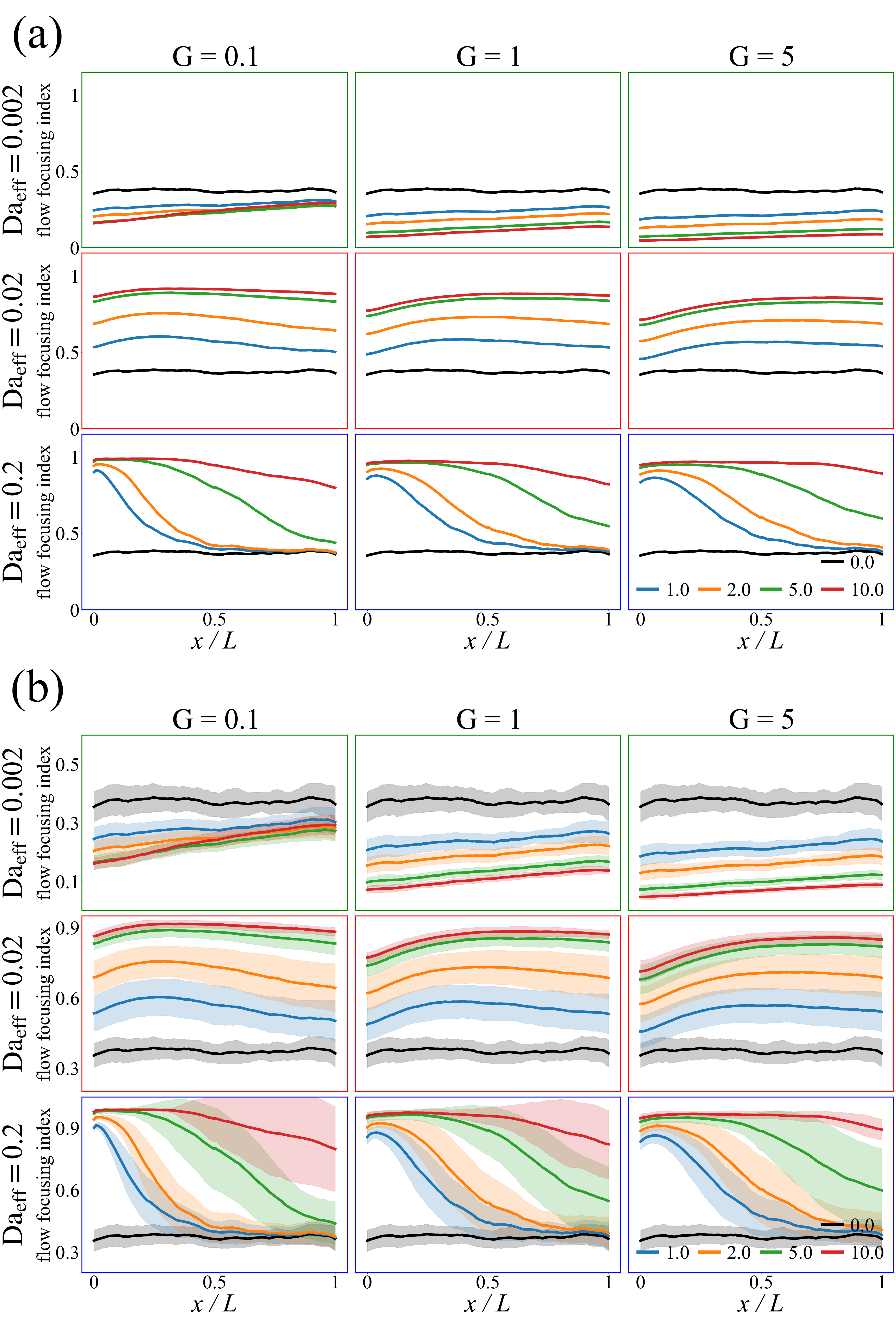}
    \caption{Evolution of the flow focusing profile for the regular pore network across the ($\textrm{Da}_\textrm{eff}$, $\textrm{G}$) parameter space. (a) Mean profile value at successive time snapshots (solid lines). (b) Same data, with ensemble variability: the shaded band around each curve denotes the mean $\pm$ 1 standard deviation across all realizations. The plots show the initial flow focusing profile (black line) and profiles at times $T = 1.0, 2.0, 5.0,$ and $10.0$ (colored lines). The frame color denotes the dissolution regime: green for uniform, red for channeling, and blue for wormholing.}
    \label{fig:regular_p}
\end{figure}

We simulate the regular pore network for $\textrm{Da}_\textrm{eff} \in \{0.002, 0.02, 0.2\}$ and $\textrm{G} \in \{0.1, 1, 5\}$. The flow focusing profile is recorded at $T \in \{0.0, 1.0, 2.0, 5.0, 10.0\}$, giving a complete history of the dissolution process. Fig.~\ref{fig:regular_r} presents the evolution of the network in the three dissolution regimes and Fig.~\ref{fig:regular_p} shows the profile evolution for each $(\textrm{Da}_\textrm{eff},\textrm{G})$ pair: panel (a) displays the full range to reveal overall trends, whereas panel (b) focuses on the range of evolution of the profile, containing additional information on variability within the ensemble.

In the uniform regime (top rows of Fig.~\ref{fig:regular_p}a,b), the flow focusing profile declines with time. The drop is steepest at high G: diffusion hindrance suppresses the growth of the widest channels, allowing narrower ones to dissolve faster so the network approaches complete homogeneity (profile values tend toward zero at $\textrm{G}=5$, as shown in the top-right panels of Fig.~\ref{fig:regular_p}a,b). At low G the values of the profile also decrease, but now because all channels enlarge at nearly the same rate, reducing their conductance contrasts. A small inlet–outlet asymmetry persists, indicating that the reactant penetration length---though large---is still finite relative to the system length.

In the channeling regime (middle rows of Fig.~\ref{fig:regular_p}a,b), profile values increase in time along the entire system length. Initial main flow paths widen almost uniformly from inlet to outlet, although the region near the inlet remains slightly more homogeneous, as more dissolution occurs there before the main flow path dominates the system. The effect of G is subtler. At low G the focusing weakens with depth, whereas at high G this decline nearly disappears, because diffusion hindrance allows more reactant to reach the deeper portions of the network. Conversely, at higher G the homogenization near the inlet is more pronounced, as dissolution in smaller channels is then faster. This inlet-outlet asymmetry is clearly visible in the dissolution pattern in Fig.~\ref{fig:regular_r}b.

In the wormholing regime (bottom rows of Fig.~\ref{fig:regular_p}a,b), a single highly conductive channel advances through the lattice, producing a sigmoidal profile: near-maximum values behind the tip and unchanged values ahead of it. The progress of the wormhole is faster for larger G---again, due to diffusion hindrance, more reactant is pushed toward the tip of the wormhole, instead of being used for widening the already existing parts of the channel.

Fig.~\ref{fig:regular_p}b shows the variability of the flow focusing profile within the ensemble. The data indicate that, although different network realizations—which vary in their initial pore-diameter distributions—evolve at different rates, the variance becomes negligible by the end of the simulation. This convergence implies that the ultimate outcome—whether the degree of homogenization in the uniform regime or a characteristic level of channelization—does not depend on the specific initial diameter distribution.

\subsection{Disordered Pore Network}
\begin{figure}
\centering
\noindent\includegraphics[width=1.0\textwidth]{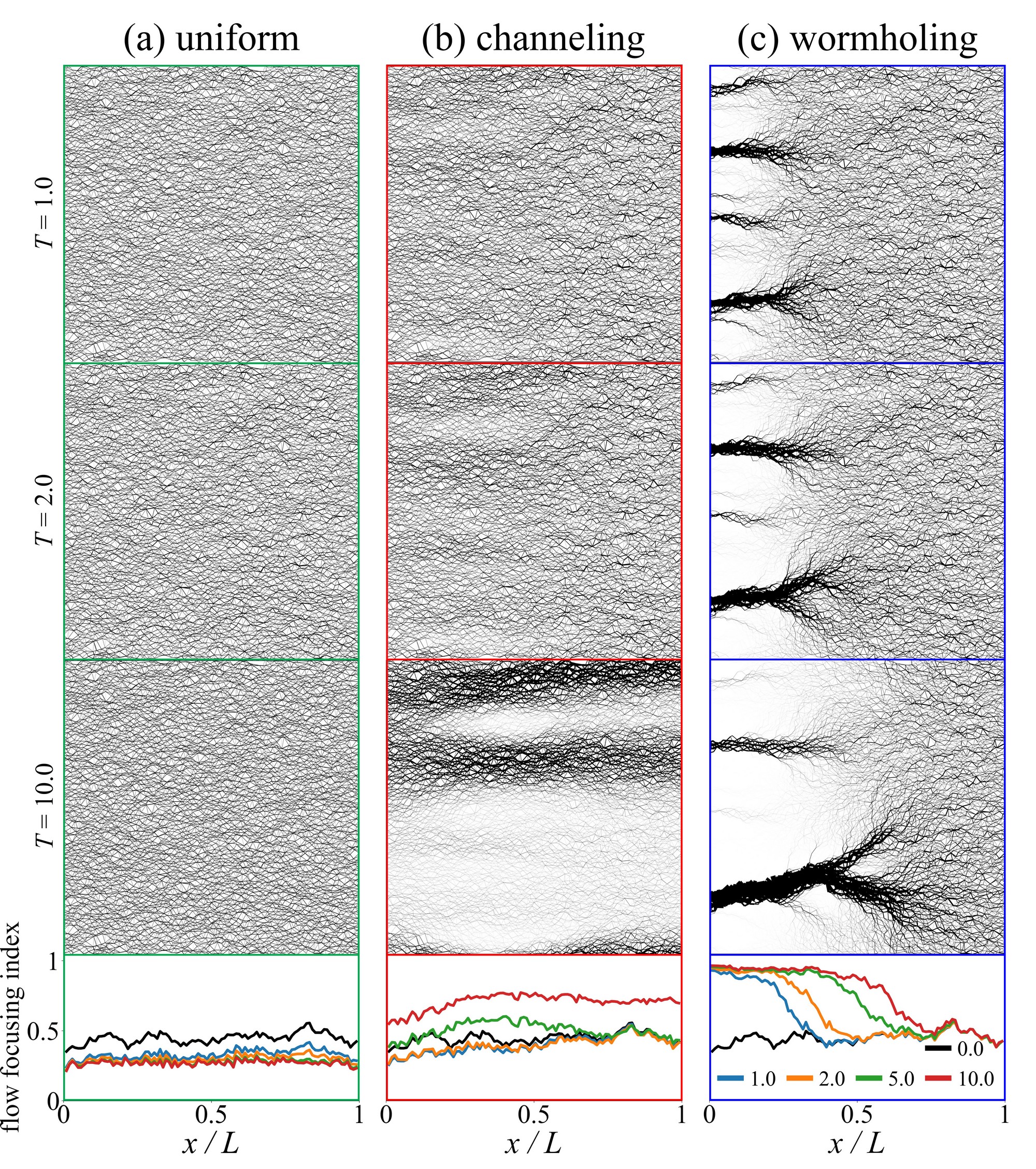}
    \caption{Evolution of the disordered pore network and the flow focusing profile in three dissolution regimes: (a) uniform ($\textrm{Da}_\textrm{eff} = 0.002$, $\textrm{G} = 5$), (b) channeling ($\textrm{Da}_\textrm{eff} = 0.02$, $\textrm{G} = 5$), and (c) wormholing ($\textrm{Da}_\textrm{eff} = 0.2$, $\textrm{G} = 5$). Edge width is proportional to the volumetric flow rate, with the same proportionality constant used in all panels. The plots show the initial flow focusing profile (black line) and profiles at times $T = 1.0, 2.0, 5.0,$ and $10.0$ (colored lines). The frame color denotes the dissolution regime: green for uniform, red for channeling, and blue for wormholing.}
    \label{fig:disordered_r}
\end{figure}

\begin{figure}
\centering
\noindent\includegraphics[width=0.80\textwidth]{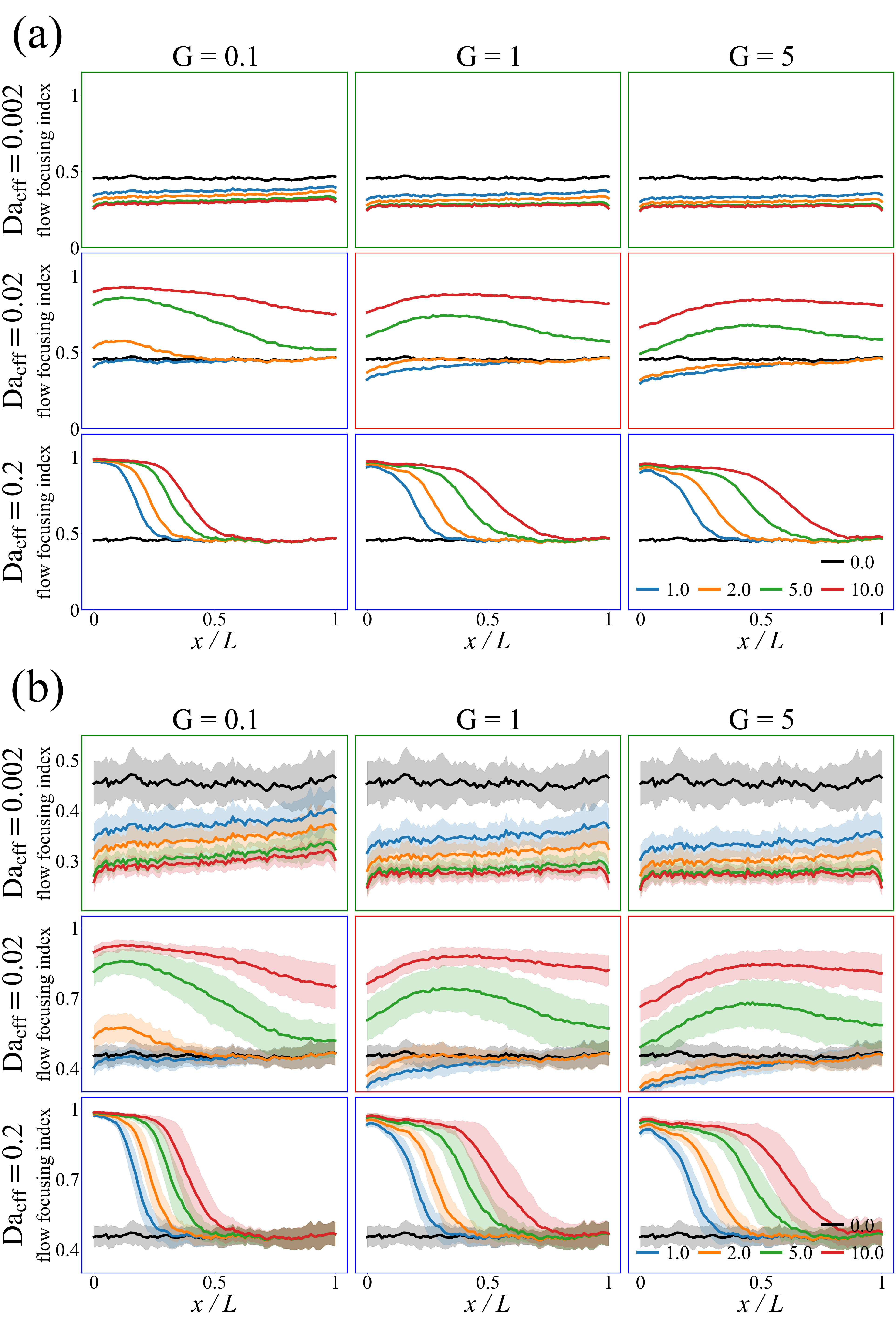}
    \caption{Evolution of the flow focusing profile for the disordered pore network across the ($\textrm{Da}_\textrm{eff}$, $\textrm{G}$) parameter space. (a) Mean profile value at successive time snapshots (solid lines). (b) Same data, with ensemble variability: the shaded band around each curve denotes the mean $\pm$ 1 standard deviation across all realizations. The plots show the initial flow focusing profile (black line) and profiles at times $T = 1.0, 2.0, 5.0,$ and $10.0$ (colored lines). The frame color denotes the dissolution regime: green for uniform, red for channeling, and blue for wormholing.}
    \label{fig:disordered_p}
\end{figure}
The results of dissolution for the disordered pore network are depicted in Figs.~\ref{fig:disordered_r} and~\ref{fig:disordered_p}. They are in agreement with our previous work.~\cite{szawello2024} We run the same set of simulations as for the regular lattice, using identical dimensionless parameters. We utilize the same diameter distributions, yet the initial flow focusing index in the ensemble is higher ($f_{50\%} \approx 0.5$ in~Fig.~\ref{fig:disordered_p} versus 0.4 in Fig.~\ref{fig:regular_p}). This difference reflects the additional segment-scale heterogeneity introduced by non-uniform pore lengths.

% The speed of evolution of the dissolution patterns highlights important contrasts between the disordered and regular pore networks. In the channeling and wormholing regimes the disordered network evolves more slowly: at the same simulation time $T$ its channels have advanced noticeably less than those in the regular lattice. By contrast, in the uniform regime the disordered lattice dissolves at a similar rate---and at high G even faster.

The dissolution patterns evolve at different rates in the disordered and regular pore networks, as shown by the patterns in Figs.~\ref{fig:regular_r} and~\ref{fig:disordered_r}. In the channeling and wormholing regimes, the disordered network advances more slowly: at the same nondimensional time $T$, the channels in the disordered network extend noticeably less than those in the regular lattice. In the uniform regime, however, the disordered lattice dissolves at a comparable rate---and, at high G, even faster—than the regular network.

Several factors contribute to the contrasting evolution of the two networks. Geometrically, the disordered lattice contains numerous edges perpendicular to the main pressure gradient.  These cross-flow channels are nearly inactive in the uniform regime, where transverse pressure differences are negligible, but they start to carry substantial flow once wormholes form, because the wormhole tips impose strong lateral pressure gradients. This is increasing the reactant demand and slowing the growth. Second, pronounced aperture and length heterogeneity promotes branching, which further retards wormhole advance.~\cite{cheung2002,upadhyay2015} When heterogeneity becomes extreme, however, stronger flow focusing can compensate for the branching penalty and accelerate growth again.~\cite{cheung2002} Because our disordered network displays significant heterogeneity at both conduit and segment scales, pinpointing its exact position on this spectrum—and how these competing effects balance—remains challenging.

The two networks differ not only in how quickly they evolve but also in how completely they can homogenize. At low Damk\"ohler numbers (top rows of Fig.~\ref{fig:disordered_p}a,b), the disordered network still enters the uniform dissolution regime, yet the resulting homogenization is weaker than in the regular network. Structural heterogeneity---independent of diameter variations---persists, and even at high G the flow focusing index plateaus at $f_{50\%}\approx 0.25$, as shown in the top-right panels of Fig.~\ref{fig:disordered_p}a,b. This saturation indicates an intrinsic limit to homogenization in the disordered medium. The snapshots in Fig.~\ref{fig:disordered_r}a change only marginally over time, confirming that much of the initial flow distribution is related to the disordered geometry of the network. This finding is crucial for studies aiming at complete homogenization: although uniform dissolution can eliminate all diameter-based heterogeneity in regular lattices,~\cite{roded2020, deng2025} it cannot remove heterogeneity rooted in the network topology. Because natural rocks are never perfectly regular---both pore diameters and lengths vary---results derived from regular pore networks cannot be generalized directly to real systems.

% At intermediate $\textrm{Da}_\textrm{eff}$ (middle rows of Fig.~\ref{fig:disordered_p}a,b), the disordered network behaves differently from the regular lattice. Its extra heterogeneity---especially the variation in pore lengths---impedes the formation of dominant flow paths spanning the entire system. Fig.~\ref{fig:disordered_r} shows that preferential channels exist from the outset, but they are far less pronounced than in the regular network. This is because the pore-diameter field is spatially correlated and thus encourages channel formation. Clusters of large-diameter pores create contiguous low-resistance corridors, so flow concentrates along them and dissolution amplifies the contrast, sharpening the channel. By contrast, the uncorrelated distribution of pore lengths work against sustained pathway development. As a result, the regime depends strongly on G, as demonstrated by the behavior of the flow focusing profile in Fig.~\ref{fig:disordered_p}. At small G, the network displays the features of the wormholing regime: a sigmoidal profile and a dominant pathway that grows gradually from inlet to outlet. At higher G, the profile resembles that of the regular network, but with more initial homogenization. Because channel growth is slower, dissolution has more time to smooth out inlet-region heterogeneity before a single path takes over.
At intermediate $\textrm{Da}_\textrm{eff}$ (middle rows of Fig.~\ref{fig:disordered_p}a,b), the disordered network behaves differently from the regular lattice. The additional heterogeneity—especially the variation in pore lengths—makes it harder for a single dominant flow path to form across the entire system. As shown in Fig.~\ref{fig:disordered_r}, preferential channels are present from the beginning, but they are much less pronounced than in the regular network. This reflects the different roles of the two types of heterogeneity. The pore-diameter field is spatially correlated, so clusters of large pores tend to line up and create continuous low-resistance corridors; flow concentrates along these corridors and dissolution then sharpens them into channels. The pore-length distribution, by contrast, is uncorrelated and therefore acts against the formation of such system-spanning paths. As a result, the dissolution regime depends strongly on G, as demonstrated by the behavior of the flow focusing profile in Fig.~\ref{fig:disordered_p}. At small G, the network displays the features of the wormholing regime: a sigmoidal profile and a dominant path that grows gradually from inlet to outlet. At higher G, the profile resembles that of the regular network, but with more initial homogenization. Because channel growth is slower, dissolution has more time to smooth out inlet-region heterogeneity before a single path takes over.

At high $\textrm{Da}_\textrm{eff}$, both networks consistently develop wormholes and their growth accelerates with increasing~$\textrm{G}$. The key difference lies in how strongly the wormholes compete. In the regular lattice, screening is stronger, so one wormhole soon outcompetes its neighbors and captures nearly the entire discharge. Once this conduit monopolizes the flow, the high velocity within it increases the reactant penetration length, broadening the transition zone in the flow focusing profile (bottom rows of Fig.~\ref{fig:regular_p}a,b). In the disordered network, weaker screening allows several channels to share the flux; the dominant wormhole advances more slowly, and the focusing profile remains steeper (bottom rows of Fig.~\ref{fig:disordered_p}a,b).

The ensemble variability shown in Fig.~\ref{fig:disordered_p}b is of the same order of magnitude as in the regular network, but—because it now reflects spatially correlated diameter heterogeneity and the randomness of node positions—the profile is less smooth. How this variability evolves depends on the dissolution regime. In the uniform regime it decreases as conduit-scale heterogeneity is removed. In the channeling regime it first increases and then declines, indicating different evolution rates for individual realizations that eventually converge to a common outcome. In the wormholing regime variable growth rates are also evident; however, because the simulation ends before wormhole breakthrough, the expected late-time reduction in variability is not yet observed.

%However, the wormholes advance more slowly and exert stronger control on the network---the focusing index is higher and the profile front steeper. A regular lattice supports more branching competition, so its wormhole tip is fed by several parallel paths, producing a broader transition zone in the flow focusing profile.

\subsection{Discrete Fracture Network}
\begin{figure}
\centering
\noindent\includegraphics[width=1.0\textwidth]{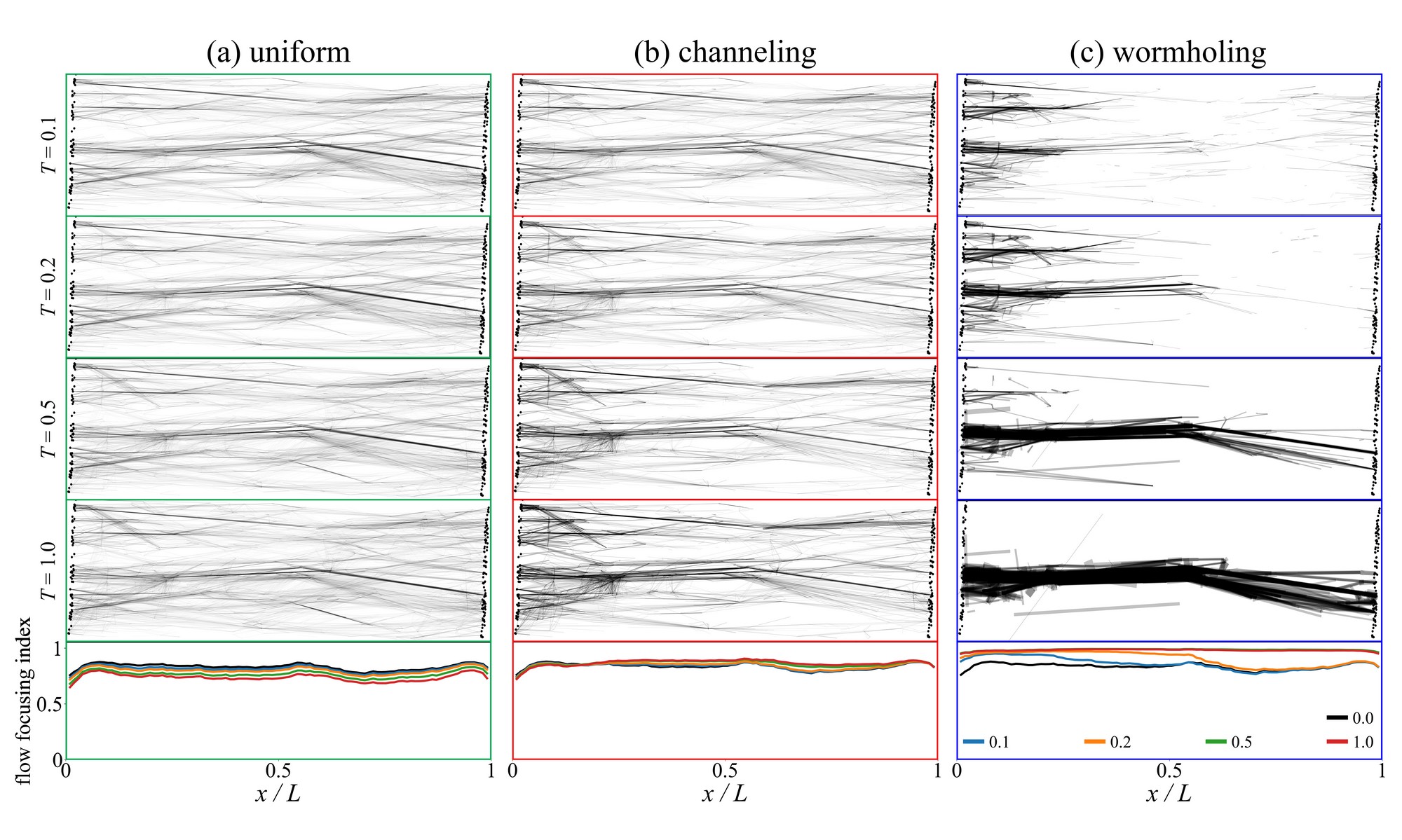}
    \caption{Evolution of the discrete fracture network and the flow focusing profile in three dissolution regimes: (a) uniform ($\textrm{Da}_\textrm{eff} = 0.0002$, $\textrm{G} = 5$), (b) channeling ($\textrm{Da}_\textrm{eff} = 0.002$, $\textrm{G} = 5$), and (c) wormholing ($\textrm{Da}_\textrm{eff} = 0.02$, $\textrm{G} = 5$). The network is shown as a graph after a principal-component-analysis projection; only edges carrying more than 1\% of the maximum flow rate in the system are plotted. Edge width is proportional to the volumetric flow rate, with the proportionality constant rescaled to maximize visibility but kept fixed within each regime. The plots show the initial flow focusing profile (black line) and profiles at times $T = 0.1, 0.2, 0.5,$ and $1.0$ (colored lines). The frame color denotes the dissolution regime: green for uniform, red for channeling, and blue for wormholing.}
    \label{fig:discrete_r}
\end{figure}

\begin{figure}
\centering
\noindent\includegraphics[width=0.80\textwidth]{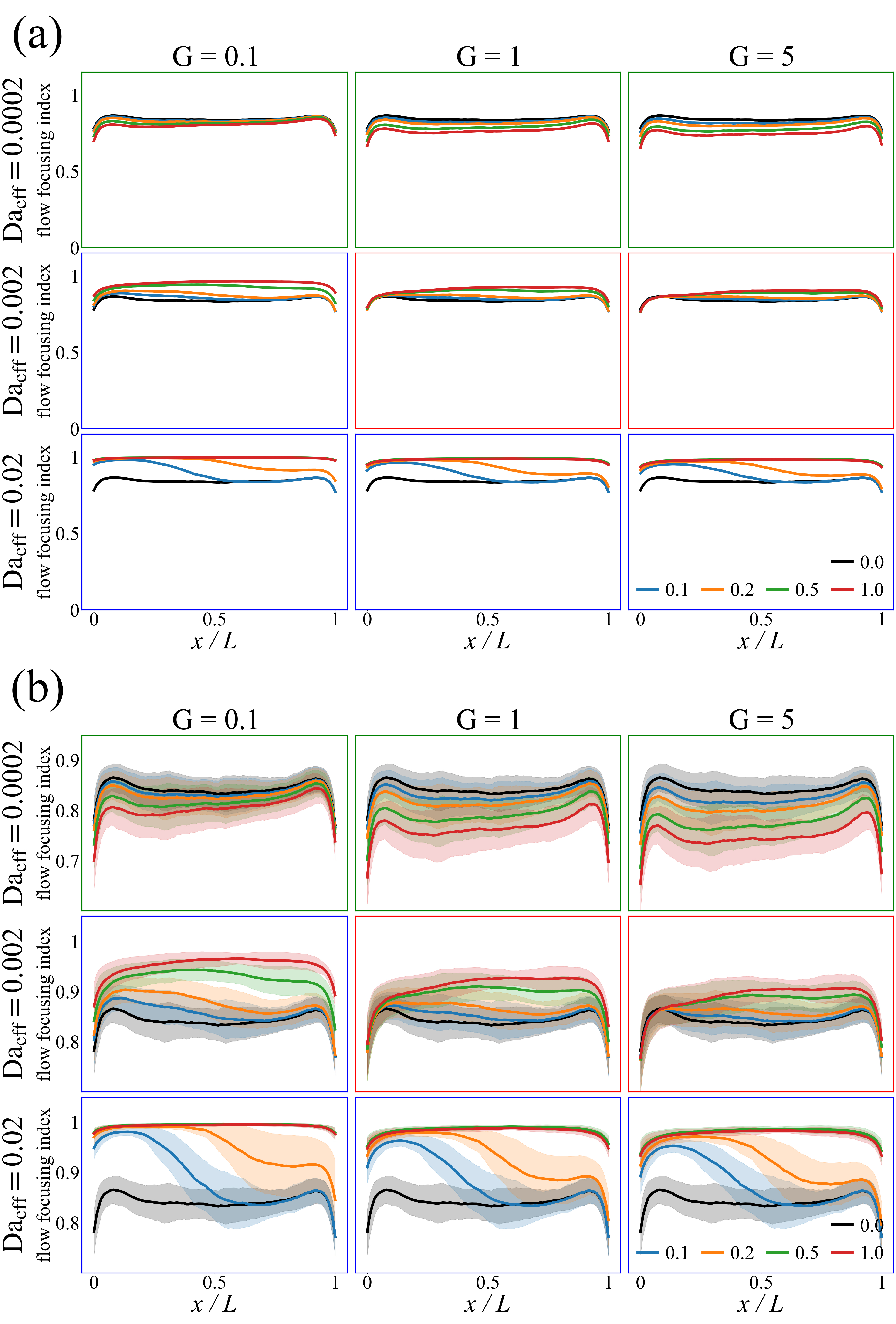}
    \caption{Evolution of the flow focusing profile for the discrete fracture network across the ($\textrm{Da}_\textrm{eff}$, $\textrm{G}$) parameter space. (a) Mean profile value at successive time snapshots (solid lines). (b) Same data, with ensemble variability: the shaded band around each curve denotes the mean $\pm$ 1 standard deviation across all realizations. The plots show the initial flow focusing profile (black line) and profiles at times $T = 0.1, 0.2, 0.5,$ and $1.0$ (colored lines). The frame color denotes the dissolution regime: green for uniform, red for channeling, and blue for wormholing.}
    \label{fig:discrete_p}
\end{figure}

Fig.~\ref{fig:discrete_r} presents the evolution of the dissolving DFNs and Fig.~\ref{fig:discrete_p} depicts the corresponding changes in the flow focusing profile across the parameter space. We perform the simulations for $\textrm{Da}_\textrm{eff} \in \{0.0002, 0.002, 0.02\}$ and $\textrm{G} \in \{0.1, 1, 5\}$, and capture data at $T \in \{0.0, 0.1, 0.2, 0.5, 1.0\}$. The initial values of the flow focusing index are very high ($f_{50\%} \approx 0.8$ in Fig.~\ref{fig:discrete_p}), indicating that the network is strongly channelized even before any reaction occurs. Because the discrete fracture network has a much larger hydraulic cross-section than the pore lattices, the effective Damköhler numbers required to traverse the three dissolution regimes are shifted roughly one order of magnitude lower. After accounting for this shift, the expected progression—uniform dissolution, channeling, and finally wormholing—emerges unchanged.
% Additionally, in fracture networks a much shorter time is required to reach a steady level of flow focusing than in porous media. This again reflects the strong pre-existing heterogeneity: flow pathways are not newly carved out, but are only slightly enhanced or slightly attenuated. Variability in the distribution and connectivity of flow paths therefore dominates over the comparatively small changes produced by aperture dissolution, so only a short dissolution time is needed for the influence of the aperture distribution to reach its maximum.

As visible in the top rows of Fig.~\ref{fig:discrete_p}a,b, in the uniform regime the flow focusing profile shows a modest decrease, maximally to around $f_{50\%} \approx 0.7$ in the top-right panels of Fig.~\ref{fig:discrete_p}a,b, but the values remain high, showing that uniform aperture growth cannot eliminate heterogeneity rooted in fracture connectivity and path lengths, making complete homogenization unattainable in a DFN.

In the channeling regime (middle and middle-right panels of Fig.~\ref{fig:discrete_p}a,b), the flow focusing profile rises only slightly with time, as expected for a network that is already strongly channelized. Its shape, however, evolves much like in porous media: the increase is nearly uniform across the entire domain. In addition, for $\textrm{Da}_\textrm{eff}=0.002$ and $\textrm{G}=0.1$, the DFN transitions to the wormholing regime, mirroring the behavior of the disordered pore network and further underscoring the influence of additional heterogeneity.

In the wormholing regime (middle-left panels and bottom rows of Fig.~\ref{fig:discrete_p}a,b), the flow focusing profile develops a transition zone, much like in porous media. A notable difference, however, is that a higher G slows wormhole advance—the opposite of the behavior seen in pore networks. At the same time, the flow focusing index no longer approaches its maximum when G is large. This indicates that diffusion limitation becomes significant, restricting dissolution in the main channel. As a result, competition among flow paths weakens; smaller channels continue to carry a portion of the flux, thereby reducing the growth rate of the dominant path.

The variability in the flow focusing index (Fig.~\ref{fig:discrete_p}b) mirrors that of the disordered pore network, rising at first and then declining as dissolution proceeds. In the DFNs the wormholes ultimately reach the outlet, and, once breakthrough occurs, the profile exhibits the expected reduction in variability.

\subsection{Homogenization by Dissolution}
\begin{figure}
    \centering
\noindent\includegraphics[width=1.0\textwidth]{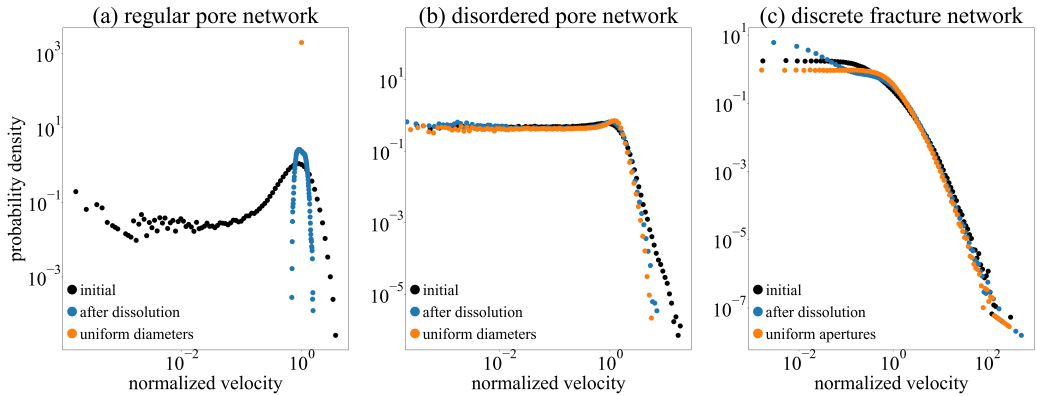}
     \caption{Velocity distributions for three network types: (a) regular pore network, (b) disordered pore network, and (c) discrete fracture network. Each panel shows three histograms: the initial distribution (black), the distribution after uniform dissolution (blue), and the distribution for the same ensemble but with uniform diameters or apertures (orange). Velocities are normalized by the ensemble-average velocity in each case.}
     \label{fig:velocity}
\end{figure}
The flow focusing data show that perfect homogenization is unattainable in either disordered porous media or DFNs. To examine why, we compare the velocity distributions for all three network types. Specifically, we sample the initial velocity distribution, the distribution after uniform dissolution (final snapshots for $\textrm{Da}_\textrm{eff}=0.002$, $\textrm{G} = 5$ in Figs.~\ref{fig:regular_p} and~\ref{fig:disordered_p}, and $\textrm{Da}_\textrm{eff}=0.0002$, $\textrm{G} = 5$ in Fig.~\ref{fig:discrete_p}), and the distribution for an ideal network with perfectly uniform diameters and apertures.

Fig.~\ref{fig:velocity} presents the comparison of these distributions. For the regular pore network, the initial histogram is broad because of diameter variability; after dissolution it narrows sharply and collapses to a single value in the uniform diameter limit. This behavior is consistent with other studies that use the regular pore network.~\cite{roded2020, deng2025} 

In the disordered network the initial velocity histogram is broad, with a high-velocity tail generated by the largest pores. Uniform dissolution trims this tail by reducing the contrast between large and small pores. At the same time, very low velocities become more common: as heterogeneity diminishes, flow aligns with the main pressure gradient and many pores oriented perpendicular to it carry negligible flux. Interestingly, the histogram after dissolution almost coincides with that of the uniform‐diameter lattice, confirming that the remaining heterogeneity---and thus the residual flow focusing---is rooted in network structure, not in diameter variance.

The DFN follows the same overall trend: the high-velocity tail shortens, and the number of low-velocity fractures increases. After dissolution the tail closely matches the uniform-aperture case, yet the low-velocity end still differs markedly. This mismatch likely results from the three fracture families in the structure of the DFN, which begin with distinct initial apertures. Dissolution broadens the aperture distribution of each family, but not enough for fractures from different families to overlap in large numbers. Forcing all apertures to a single initial value would eliminate this discrepancy—but at the cost of fundamentally altering the structure of the network.

Whether we look at the network after dissolution or at an idealized version with uniform diameters or apertures, the velocity distributions are still far from the delta-like peak achieved by the regular pore lattice. All residual variance is therefore dictated by structural heterogeneity. Because that heterogeneity is greater in the DFN than in the disordered pore network, the lowest attainable flow focusing index is likewise higher. The principle is the same in every case: once the network geometry sets a structural limit, additional dissolution cannot homogenize the flow any further.

\section{Conclusions}
Dissolution reshapes both structure and flow in porous and fractured media---systems whose contrasting architectures and transport behaviors influence many geologic and engineering processes. To quantify these effects, we perform network-based simulations on three models: a regular pore network, a disordered pore network, and a discrete fracture network, spanning a wide range of flow and reaction conditions. 
The results are quantified using the flow focusing profile. Every dissolution regime presents distinct characteristics in each system. In the uniform regime, all networks are homogenized and flow focusing decreases across the entire domain. However, while the flow in the regular medium becomes completely uniform, the disordered medium retains a certain level of focusing, and the discrete fracture network remains highly heterogeneous, with only minimal redirection of flow into parts of the network beyond the main paths. 

In the channeling regime, we observe an increase in flow focusing along the whole system, with slight inlet–outlet asymmetry in the regular medium and a much more pronounced asymmetry—with significant homogenization near the inlet—in the disordered medium. For the discrete fracture network, this increase is minimal, as the flow is highly channelized from the outset. In the wormholing regime, a front in the flow focusing profile progresses from the inlet, marking wormhole formation, but the speed of the wormhole and the sharpness of the front differ between the systems.
 
We focus our analyses on the interplay between heterogeneity and uniform dissolution. The three systems exhibit different types and degrees of heterogeneity, and each responds to dissolution in its own way. In the regular network, heterogeneity arises solely from the pore-diameter distribution and is readily eliminated by uniform dissolution. In the disordered network and the DFN, additional structural heterogeneity persists: the lengths of individual paths and the connectivity between parts of the network remain unchanged. Uniform dissolution therefore removes only the heterogeneity associated with diameter or aperture variations, demonstrating that the inherent structure exerts strong control over dissolution dynamics.

These findings carry important implications for continuum-scale modeling. In our simulations, dissolution can eradicate heterogeneity in conduit diameters (or apertures) but cannot remove heterogeneity rooted in path lengths and network connectivity; as a result, flow never becomes fully uniform in the disordered lattice or in the DFN. Darcy-scale finite-difference models, however, represent the medium only through spatially varying porosity or permeability fields. Because those fields evolve toward uniform values when diameters enlarge uniformly, such models would predict nearly complete flow homogenization under uniform dissolution—an outcome that, according to our network results, is unattainable in natural media where connectivity and length distributions persist. Capturing the true limits of homogenization will therefore require numerical approaches that retain at least some information on network topology, for example through higher-order upscaling techniques or hybrid continuum–network formulations.

Our work extends the application of the flow focusing profile from disordered pore networks~\cite{szawello2024} to a broader class of systems, including regular networks and discrete fracture networks, and shows that conduit-, segment-, and network-scale heterogeneity each leave a distinct imprint on dissolution regimes. The metric provides a unified way to compare porous and fractured media across parameter space and to identify when channeling, wormholing, or uniform dissolution are limited by quenched structural disorder. The strongly varying levels of this disorder in porous media and fracture networks indicate that care is needed when extrapolating laboratory experiments—often based on core dissolution in porous media—to field-scale fracture networks. The emergent properties of the dissolving medium, and its transport behavior, may differ substantially from those inferred from such experiments. This implies that predictions of stimulation efficiency, injectivity evolution, or reactive-front propagation in applications such as acidizing, geothermal operations, and geological carbon storage must account for the structural heterogeneity encoded in path lengths and connectivity. Designs or models that assume uniform dissolution can fully homogenize flow are likely to underestimate the persistence of preferential paths, and therefore the degree of channelized transport, even in regimes that appear uniform at the pore scale.

% Although we analyze porous media and fracture networks separately, natural formations often combine both, complicating the picture.~\cite{berre2019} Our results indicate that porous matrix dissolution can differ markedly from fracture-controlled dissolution. Future work should therefore examine coupled processes, with many subtle interactions to consider---such as dissolution-induced fracturing~\cite{shovkun2019} or matrix fragments clogging fractures~\cite{ellis2013}---and determine how the couplings between porous matrix and fractures affect the overall dissolution behavior.

\section{Data Availability}
All simulation data files are available in [dataset] Szawełło~et~al.~\cite{szawello2025_data} The software for dissolution and flow focusing profile calculation is provided in Szawełło.~\cite{szawello2025_soft}

\section{Acknowledgements}
Research of J. D. Hyman and P. K. Kang was supported as part of the Center on Geo-process in Mineral Carbon Storage, an Energy Frontier Research Center funded by the U.S. Department of Energy, Office of Science, Basic Energy Sciences at the University of Minnesota under award \#DE-SC0023429. Research of T. Szawełło and P. Szymczak was supported by the National Science Centre (NCN; Poland) under research Grant 2022/47/B/ST3/03395. Research of J. D. Hyman was also partially supported by CUSSP (Center for Understanding Subsurface Signals and Permeability),  funded by the U.S. Department of Energy (DOE), Office of Science under FWP 81834. Los Alamos National Laboratory is operated by Triad National Security, LLC, for the National Nuclear Security Administration of U.S. Department of Energy (Contract No. 89233218CNA000001). LANL unclassified release number LA-UR-25-26376.

\appendix
\section{DFN Generation}
\label{appendix:dfngeneration}
We generate three-dimensional discrete fracture networks (DFN) using the {\sc dfnWorks}~\cite{hyman2015} software suite. 
The cornerstone of the DFN methodology is that fractures tend to have a radius (length) that is much larger than their aperture (height). 
This disparity is used to justify a co-dimension one representation of each fracture. 
For example, each fracture is a one-dimensional line in a two-dimensional simulation or a two-dimensional plane in a three-dimensional simulation. 
Thus, in our 3D DFN model, each fracture is represented as a rectangle using the methods described in Hyman~et~al.~\cite{hyman2014}
These planes intersect with one another to form a network, through which flow and transport is simulated. 
The size, shape, orientation, and other hydrological properties are sampled from distributions whose parameters are determined from a site characterization, cf. Viswanathan~et~al.~\cite{viswanathan2022} for a comprehensive discussion of DFN modeling approaches.
Thus, {\sc dfnWorks} stochastically generates three-dimensional DFNs with desired characteristics such as length distributions and fracture intensities. 
Details of {\sc dfnWorks} in terms of algorithms and various applications can be found in Hyman~et~al.~\cite{hyman2015}

We consider a semi-generic DFN loosely based on the fractured carbonate-hosted Pietrasecca Fault in the central Apennines, Italy, cf. Smeraglia~et~al.~\cite{smeraglia2021} for detailed information about the fractured media. 
The domain is 25 m $\times$ 10 m $\times$ 10 m. 
We adopt three families from the survey and modify a few of their properties for our semi-generic study. 
Generation parameters of the networks are provided in Table~\ref{tbl:dfn_params}.
We measure network surface area in terms of the fracture intensity [m$^{-1}$], which is the sum of surface area of each fracture ($S_f$) divided by the volume of the domain,
 \begin{equation}\label{eq:p32}
P_{32} = \frac{1}{V}\sum_f  S_f.
\end{equation}
The fracture radii of all three families are sampled from a truncated power-law distribution with decay exponent $\alpha$, minimum radius $r_0$, and maximum radius $r_u$, and probability density function 
\begin{equation}
    p_r(r,r_0,r_u)=\frac{\alpha}{r_0}\frac{(r/r_0)^{-1-\alpha}}{1-(r_u/r_0)^{-\alpha}}.
\end{equation}
We modify the minimum and maximum fracture radius from the data provided in Smeraglia~et~al.~\cite{smeraglia2021} 
This modification requires a change in the values of fracture intensity $P_{32}$.
Fracture family orientations are sampled from a three dimensional von Mises Fisher distribution,
\begin{equation}\label{eq:fisher}
f({\bf x}; {\boldsymbol \mu}, \kappa ) = \frac{ \kappa \exp( \kappa {\boldsymbol \mu}^{T} {\bf x} )}{ 4 \pi \sinh(\kappa)}.
\end{equation}
In \eqref{eq:fisher},  ${\boldsymbol \mu}$ is the mean direction vector of the fracture family, $T$ denotes transpose, and $\kappa \geq 0$ is the concentration parameter that determines the degree of clustering around the mean direction. 
Values of $\kappa$ close to zero lead to a uniform distribution of points on the sphere while larger values create points with a small deviation from mean direction.
${\bf x}$ is a random 3-dimensional unit vector within entries sampled from a uniform distribution on $[0,1]$. Trend and Plunge are converted to a 3D normal vector using the standard convention where an upward-pointing normal (negated from the standard downward pole) is given by
\begin{align}
    {\boldsymbol \mu}[0] &= -\sin(\textrm{Trend})\cos(\textrm{Plunge}), \nonumber \\
    {\boldsymbol \mu}[1] &= -\cos(\textrm{Trend})\cos(\textrm{Plunge}), \\
    {\boldsymbol \mu}[2] &= \sin(\textrm{Plunge}). \nonumber
\end{align}
To obtain uniformly random orientations, we set $\kappa = 0.1$ and mean normal vector of $(0,0,1)$.
The distribution is sampled using the method detailed in Wood.~\cite{wood1994}
 \begin{table}
\begin{tabular}{ |p{3cm}||p{3cm}|p{3cm}|p{3cm}|  }
 \hline
 \multicolumn{4}{|c|}{DFN Families} \\
 \hline
    Parameter & Family \#1 & Family \#2  & Family \#3\\
    \hline
    Aspect ratio & 2 & 2 & 2\\
    $P_{32}$ [m$^{-1}$] & 2.52  & 4.84 & 0.315  \\
    Aperture [m] & $2.23 \times 10^{-6}$ & $9.02 \times 10^{-6}$ & $3.67\times 10^{-6}$ \\
 \hline
    $\alpha$   & 2.16    & 1.31 &   2.31\\
    $r_0$ [m]   &   1  & 1   &1\\
    $r_u$  [m]  &   5  & 5   &5\\
 \hline
    $\kappa$    &   39.9  & 61.3   &4.72\\
    Trend [$^\circ$]    &   330  & 337   &263\\
    Plunge [$^\circ$]    &   86  & 87   &51\\
 \hline
\end{tabular}

 \caption{Discrete fracture network parameters used for generating various fracture families. \label{tbl:dfn_params}}
\end{table}

Fractures from each family are placed into the domain with equal probability until the target $P_{32}$ values are attained. 
After the network is generated, isolated fractures and clusters of fractures, those that do not connect inflow and outflow boundaries, are removed because they do not participate in flow and transport. 
We generate thirty independent and statistically identical networks. 

\subsection{Graph Representation}
Once the networks are generated, we derive a graph/pipe-network representation using the method presented in Hyman~et~al.~\cite{hyman2018}
At the core of the DFN methodology is the conceptual model of a set of fractures, which are discrete entities intersecting with one another to form a network.
The following maps a DFN $\mathcal{F}$ made up of a set of $n$ fractures $\{f_i\}$ to a graph $\mathcal{G}$ composed of a set of vertices $V$ and edges $E$. 
If two fractures $f_i$ and $f_j$ intersect, $ f_i \cap f_j \ne \emptyset$, then there is a vertex $v \in V$, 
\begin{equation}\label{eq:psi1}
 \psi : f_i  \cap f_j \rightarrow v
\end{equation}
 that represents the line of intersection between the fractures $f_i $ and $f_j$.
 If $ f_i \cap f_j \ne \emptyset$ and $ f_i \cap f_k \ne \emptyset$, then there is an edge in
 $E$ connecting the corresponding vertices, 
\begin{equation}\label{eq:psi2}
 \psi : f_i \cap f_j \ne \emptyset \text{ and } f_i \cap f_k \ne \emptyset \rightarrow e(u,v) \in E.
\end{equation}
Under the mapping $\psi$ each fracture is represented by a $k$-clique where $k$ is the number of intersections on the fracture. 
Thus, each edge can be thought of as residing on a single fracture and edge weights can represent hydrological and geometric properties of that fracture.

\bibliographystyle{model6-num-names}
\bibliography{references}

\end{document}